# GENERIC SYSTEM VERILOG UNIVERSAL VERIFICATION METHODOLOGY BASED REUSABLE VERIFICATION ENVIRONMENT FOR EFFICIENT VERIFICATION OF IMAGE SIGNAL PROCESSING IPS/SOCS

Abhishek Jain[1], Giuseppe Bonanno[2], Dr. Hima Gupta[3] and Ajay Goyal[4]

[1]Imaging Group, STMicroelectronics, Greater Noida, India
[1]JBS, Jaypee Institute of Information Technology (JIIT), Noida, India
`abhishek-mmc.jain@st.com;ajain_design@yahoo.co.in`
[2]Imaging Group, STMicroelectronics, Grenoble, France
`giuseppe.bonanno@st.com`
[3]JBS, Jaypee Institute of Information Technology (JIIT), Noida, India
`hima.gupta@jiit.ac.in`
[4]Cadence Design System, Noida, India
`gajay@cadence.com`

**ABSTRACT**

*In this paper, we present Generic System Verilog Universal Verification Methodology based Reusable Verification Environment for efficient verification of Image Signal Processing IP's/SoC's. With the tight schedules on all projects it is important to have a strong verification methodology which contributes to First Silicon Success. Deploy methodologies which enforce full functional coverage and verification of corner cases through pseudo random test scenarios is required. Also, standardization of verification flow is needed. Previously, inside imaging group of ST, Specman (e)/Verilog based Verification Environment for IP/Subsystem level verification and C/C++/Verilog based Directed Verification Environment for SoC Level Verification was used for Functional Verification. Different Verification Environments were used at IP level and SoC level. Different Verification/Validation Methodologies were used for SoC Verification across multiple sites. Verification teams were also looking for the ways how to catch bugs early in the design cycle? Thus, Generic System Verilog Universal Verification Methodology (UVM) based Reusable Verification Environment is required to avoid the problem of having so many methodologies and provides a standard unified solution which compiles on all tools.*

*The main aim of development of this Generic and automatic verification environment is to develop an efficient and unified verification environment (at IP/Subsystem/SoC Level) which reuses the already developed Verification components and also sequences written at IP/Subsystem level can be reused at SoC Level both with Host BFM and actual Core using Incisive Software Extension (ISX) and Virtual Register Interface (VRI)/Verification Abstraction Layer (VAL) approaches. IP-XACT based tools are used for automatically configuring the environment for various imaging IPs/SoCs. Although this paper focus on Generic System Verilog Universal Verification Methodology based reusable verification environment built for imaging IPs/SoCs. Same concept can be extended for non imaging IPs/SoCs.*

## 1. INTRODUCTION

Imaging group has mainly two types of devices - sensors and processors. The sensors main function is to convert the viewed scene into a data stream. The companion processor function is to manage the sensor so that it can produce the best possible pictures and to process the data stream into a form which is easily handled by upstream mobile baseband or MMP (Multi-Media Processor) chipsets.

Image signal processing algorithms are developed and evaluated using Python models before RTL implementation. Once the algorithm is finalized, Python models are used as a golden reference model for the IP development. To maximize re-use of design effort, the common bus protocols are defined for internal register and data transfers. A combination of such configurable image signal processing IP modules are integrated together to satisfy a wide range of complex video processing SoCs.

"e" (Specman)/Verilog based Verification Environment was used for IP/Subsystem level verification and C/C++/Verilog based Directed Verification Environment for SoC Level Verification.

Main Challenges of Previous Environment/Verification Methodology were as follows:

1. Reusability
    a. Test cases from IP/Subsystem level could not be reused at SoC Level.
2. Maintainability
    a. Different Verification Environments were used in different IP and SoCs across multiple sites.
3. Significant time was spent in reproducing the issue reported at SoC level at IP/Subsystem level.
4. How to catch bugs early in the design cycle?
5. Verification of registers at SoC level was not efficient and automatic as small change in the register description caused manual rework in the verification environment and testcase(s).
6. At SoC level, it was difficult to align design specification with corresponding RTL implementation and verification environment.

To overcome above challenges, System Verilog Universal Verification Methodology is adopted. The main aim of development of Generic System Verilog Universal Verification Methodology based reusable verification environment is to -

1) Develop standard unified methodology across all sites which is
    a) Vendor independent
    b) Reusable from IP -> Subsystem -> SoC both with Host BFM and actual Core.
    c) Open
    d) Leveraging Existing Verification Environment

2) Usage of IP-XACT based tools for automatically generating Imaging IPs/SoCs dependent files.

The following sections discuss the new verification environment.

## 2. GENERIC SYSTEM VERILOG UNIVERSAL VERIFICATION METHODOLOGY BASED REUSABLE VERIFICATION ENVIRONMENT

### 2.1 IP Level Verification Environment

IP level verification is key aspect in SoC level verification as at SoC level, each IP is a black box and is considered as a golden block. Extensive and Exhaustive IP verification is a key requirement from protocol and functionality perspective. It is imperative to verify each and every feature of IP to greatest extent possible and delivering a zero bug IP to SoC team.

In an image signal processing IP, there are *A* input video data interfaces, *C* output video data interfaces, *B* memory interfaces, *D* output Interrupts and *E* register interfaces, where *A, B, C, D* and *E* values can be from 0 to any arbitrary number.

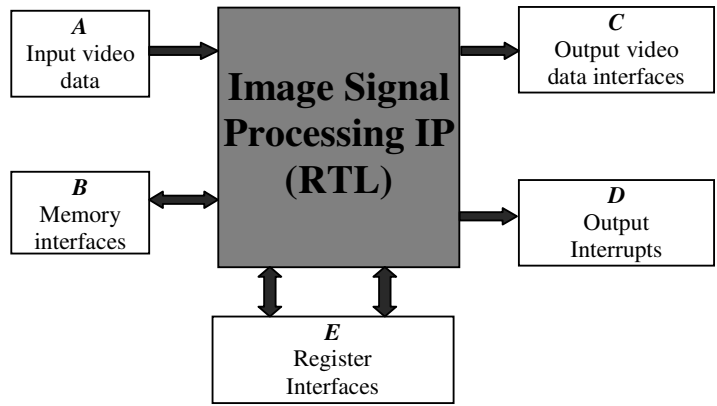

Figure. 1 Image Signal Processing IP Block Diagram

For verifying these interfaces, dedicated UVCs are used. In case of register interface(s), register interface UVC and UVM_RGM register model are used. Similarly for video data interface(s), video data interface UVC is used and for verification of interrupts, generic interrupt checker is used.

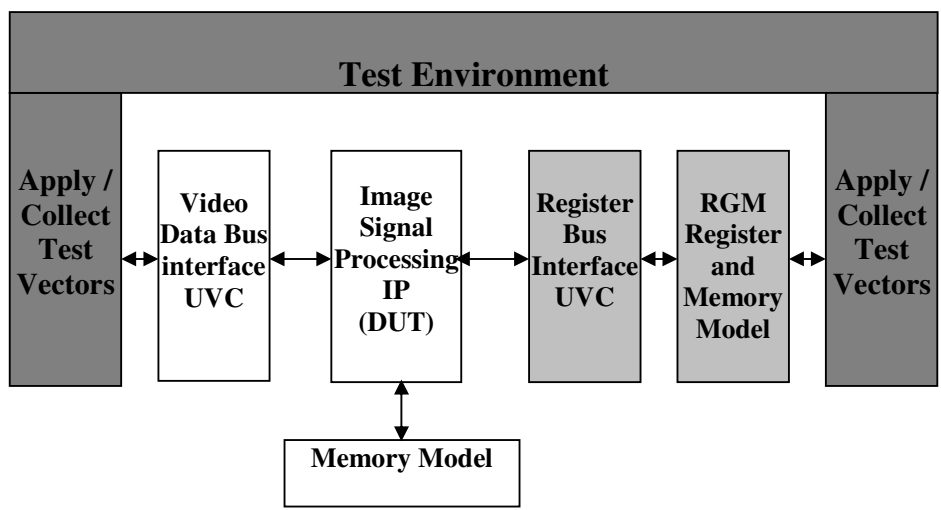

Figure. 2 Basic blocks of System Verilog UVM based IP Level Verification Environment

Note that there can be multiple instances of these UVC's in a verification environment. Each agent is configured separately and any combination of agent configurations can coexist in the same environment. Therefore in above case, *E* instances of register interface UVC agents, M (M = max (A, C)) instances of video data interface UVC agents and D instances of interrupt checker are used to interface with a DUT. Figure 2 illustrates the basic blocks of System Verilog UVM based IP Level Verification Environment.

## 2.2 Subsystem Level Verification –

At subsystem level, various IPs are connected may be via Interconnects and becomes more complex from verification perspective. It is very important to reuse the IP level verification environment to reduce the verification effort at Subsystem level.

### Reuse of Environment –

All internal IP level verification environments are configured as Passive agents whereas Interface IP level verification environments are used as Active agents. UVM-ML (Multiple Language) approach helps us in reusing the existing verification components.

### Reuse of Sequences –

**Register read/write sequences:** UVM_RGM register read/write sequences write and read address mapped registers in the DUT. As UVM_RGM have API that is independent of the bus protocol and hence can be reused at Subsystem and SoC level as at Subsystem and SoC level base address of these registers changes and in register sequences, registers can be accessed using name or type also. UVM_RGM register package is used to lookup the register address by name. UVM_RGM register package built-in sequences supports this kind of abstraction. This makes these sequences reusable and maintainable because there is no need to update the sequence each time a register address changes.

**Virtual sequences on accessible interfaces (IP3, IP4, IP5 and IP6) at subsystem-level:** These sequences are reusable from IP level to subsystem-level; some of them can be used to verify the integration of IP's into sub-system.

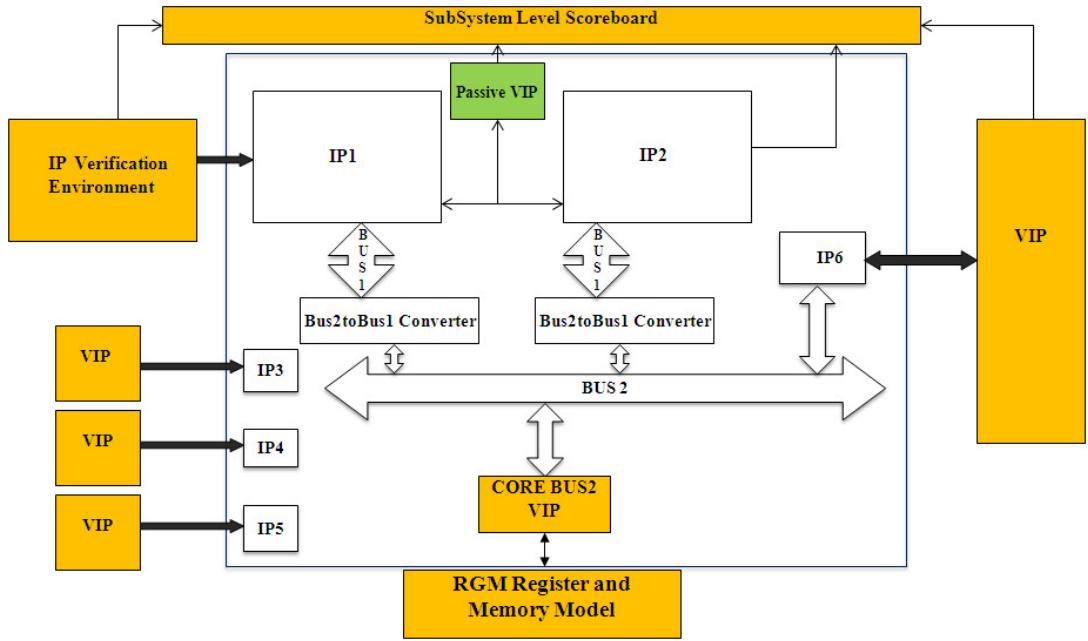

Figure. 3 Reuse of IP Level Verification Environment at Sub-System Level

### 2.3 SoC Level Verification Environment

Various subsystems are integrated together to build a SoC and make a verification task very challenging at SoC level. There are various challenges at SoC level verification like

- ➢ How to reuse subsystem level verification environment at SoC level to minimize the verification effort at SoC level

- ➢ Connectivity between IPs

- ➢ Verification of System Level scenarios

- ➢ How to synchronize "C" testcases running on Core with IP/Subsystem level verification environments to enable maximum reuse

At IP/Subsystem level verification, Cores are usually stub to do verification and BUS UVCs are used to generate BUS traffic which can be replaced by Core at SoC level. Working with a verification component at the SoC level makes it difficult to create activity that will be similar to the way the design will behave with CPU and software. An even more important challenge is what to do about SoC initialization. Sometimes, there are thousands of programmable registers that must be configured before SoC is ready to do any meaningful activity. Besides being a tedious process, the motivation for writing a long initialization sequence just for verification is low because in the end it's the job of the software to initialize the SoC, not the verification engineer. The result is duplication of effort.

International Journal of VLSI design & Communication Systems (VLSICS) Vol.3, No.6, December 2012

At SoC level, we reuse register sequences and all internal IP/Subsystem level verification environments are configured as Passive agents whereas Interface IP level verification environments are used as Active agents. Since the register sequences are independent of BUS protocol, it enables to reuse of the register sequences with different BUS at IP/Subsystem/SoC level.

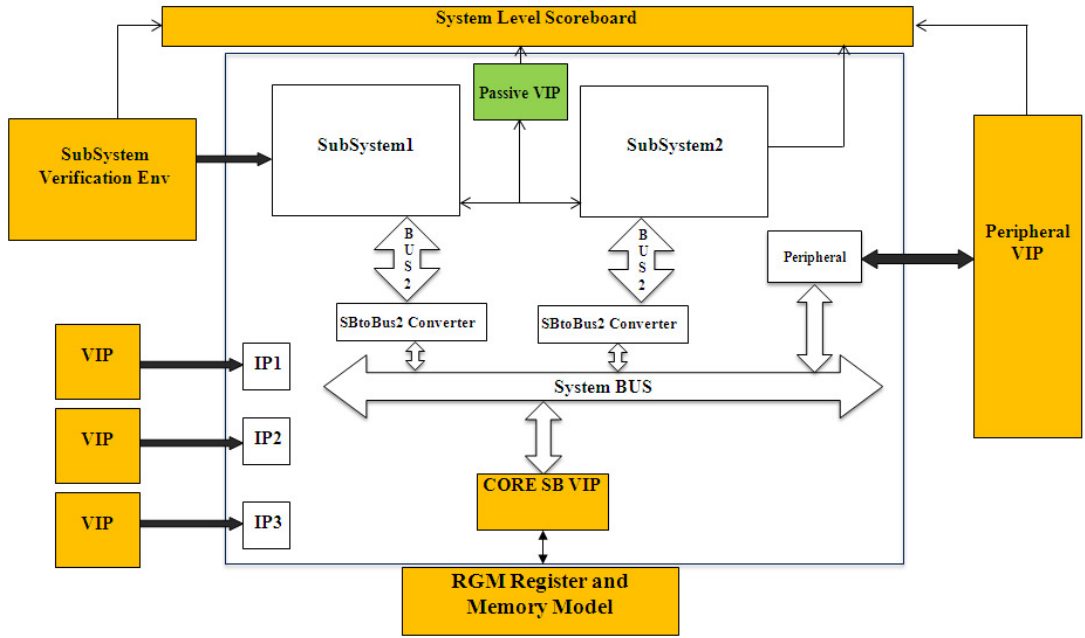

Figure. 4 Reuse of Sub-System Level Verification Environment at SoC Level

When we replace Host BFM with actual Core then it becomes challenging to reuse the existing verification environment as with Core in place, "C" testcases are used to do verification. At SoC level, it is important to verify the hardware and software works seamlessly together to deliver the functionality and performance of the system. Below are the 2 approaches which is used in imaging group -

1) We use Incisive Software Extension (ISX) to reuse the IP/Subsystem level verification environment. In this approach, Hardware/Software Co-Verification technique is used in which "C" routines are controlled/called from HVLs like "e" or "System Verilog". It enables to do constrained random and coverage driven verification of embedded software. It enable users to provide

    a. Constrained random values to "C" functions parameters
    b. Functional coverage of "C" variables
    c. Random calling of "C" functions

This helps in performing thorough verification of hardware and software together and enables in getting corner cases.

Virtual sequences from IP/Subsystem level verification environment are reused at SoC level. In this approach, "C" testcases were controlled from virtual sequences. A SW UVC was created which enabled control of "C" testcases from HVL verification environments. Since the SW UVC is in HVL, it can be used with rest of the SoC level verification environment which enables reuse of the IP/Subsystem level verification environment. At IP/Subsystem level, virtual sequences were calling register sequences over HOST BFM, which can be reused to call "C" routines which is performing register read/write to IPs via SW UVCs. SW UVC contains sequences which correspond to each "C" routine which is executing on the core. From Virtual sequences we can call these SW UVC sequences which in turn will call "C" routines.

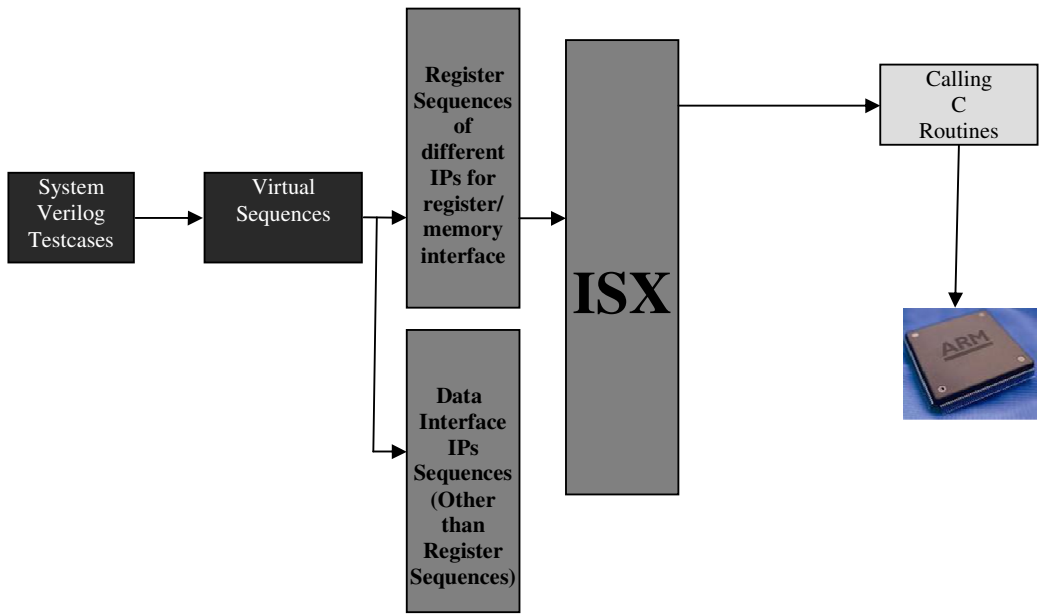

Figure. 5 Use of ISX (Incisive software extensions) in System Verilog UVM based SoC Level Verification Environment

In this approach, by connecting and controlling the C functions from the verification environments, the best of both worlds is achieved which is running C code on the CPU and the generation, checking and coverage provided by Coverage Driven Verification environment.

Incisive Software Extension is having a concept of Generic Software Adapter (GSA) which is used to connect to and control embedded software. It enables verification engineers to hook their IP/Subsystem level verification environments at SoC level executing embedded software.

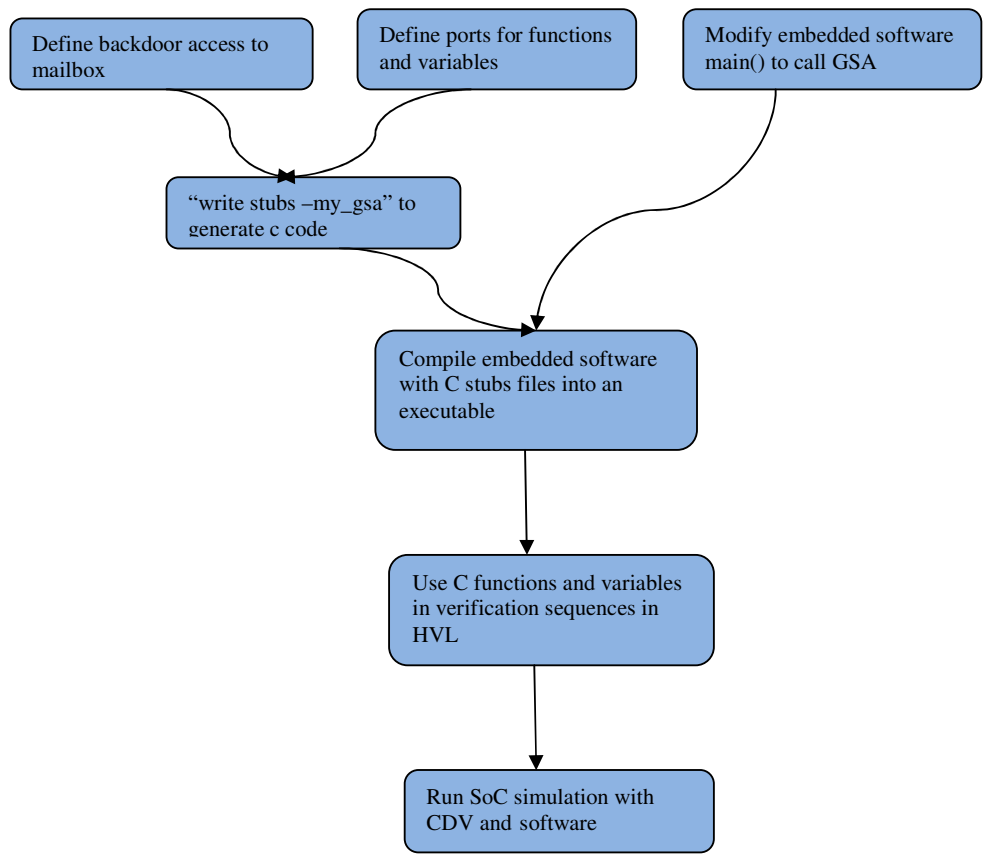

Figure 6. GSA Integration Flow. Source: Chip Design Magazine**.**

2) Second approach is to reuse IP/Subsystem verification environments from "C" testcases running on the Core. Today, most of the embedded test infrastructure uses some adhoc mechanism like "shared memory" or synchronisation mechanism for controlling simple Bus functional models (BFMs) from an embedded software.

In order to provide full controllability to the "C" test developer over these verification components, a virtual register interface layer is created over these verification environments which provides the access to the sequences of these verification environment to the embedded software enabling configuration and control of these verification environments and provide the same exhaustive verification at SoC Level.

This approach addresses the following aspects of verification at SoC Level

- ➢ Configuration and control of verification components from embedded software
- ➢ Reusability of verification environments from IP to SoC
- ➢ Enables reusability of testcases from IP to SoC
- ➢ Providing integration testcases to SoC team which is developed by IP verification teams.

It has been achieved by using Verification Abstraction Layer (VAL)/Virtual Register Interface (VRI) layer over Verification components. VAL/VRI layer over verification components is

- A virtual layer over verification environment to make it controllable from embedded software

- Provides high level C APIs hiding low level implementation

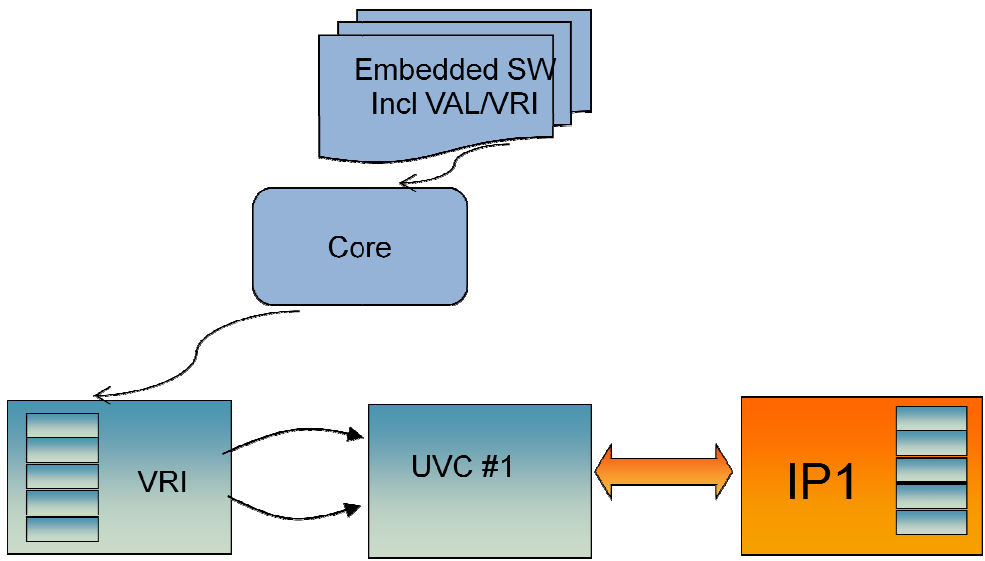

Figure 7. VAL/VRI layer over UVC used for IP verification

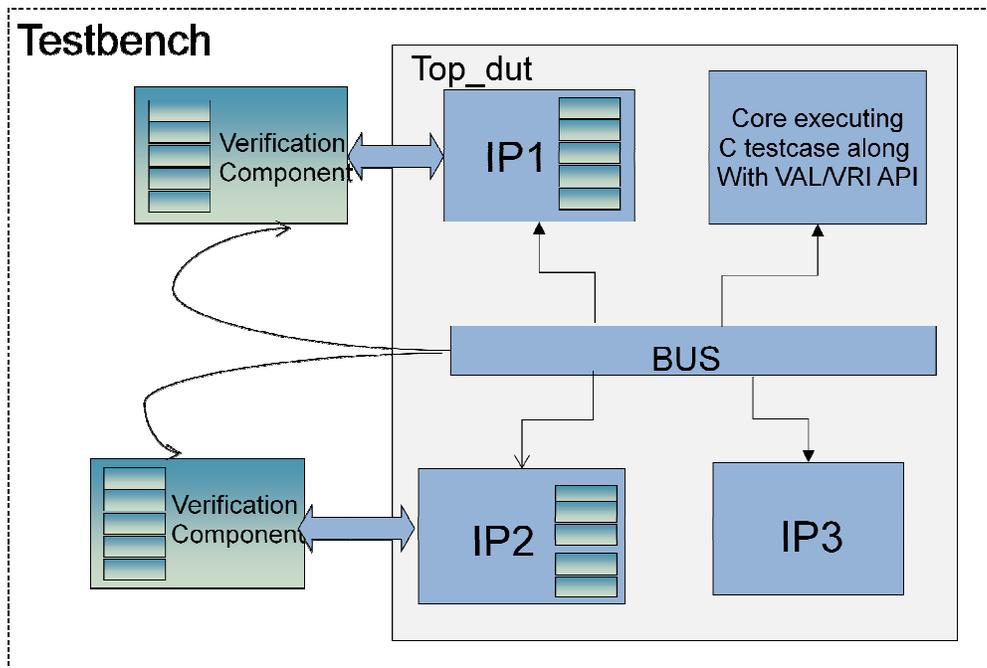

Figure 8. Reusing the IP level verification environment at SoC Level (using VAL/VRI layer)

21

## 2.4 IP-XACT Flow

In System Verilog UVM based Verification Environment, register description file for register model, address map file, sequences file, functional Coverage file, data checker file to compare the output of RTL with output of Reference(Python) model are IP/SoC specific which need to be modified for every IP/SoC. Therefore, IP-XACT based tools are used for generation of these files. First, the register map description has to be provided in XML-based IP-XACT view.

XML-based IP-XACT view is automatically generated from the Register Specification Document.

In Data checker file, there is invocation of executable of Python model containing attributes thus, automatic generation of data checker file requires the mapping between the registers/register-fields/parameters of RTL and the attributes of Python model. Thus, in IP-XACT based tools, there are two input files

1. IP-XACT view of register map, containing the register description.

2. Map file for mapping between the registers/register-fields/parameters of RTL and the attributes of Python model.

IP-XACT based tools generates IP/SoC specific files which are used in the System Verilog UVM based verification environment as shown in figure 9.

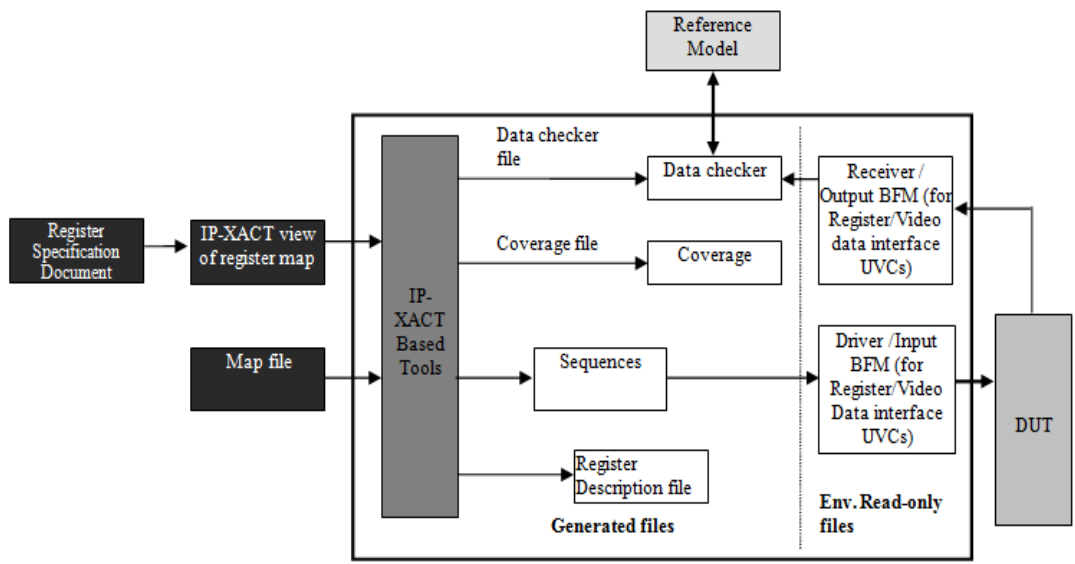

Figure. 9 IP-XACT Flow usage in System Verilog Universal Verification Methodology based Verification Environment

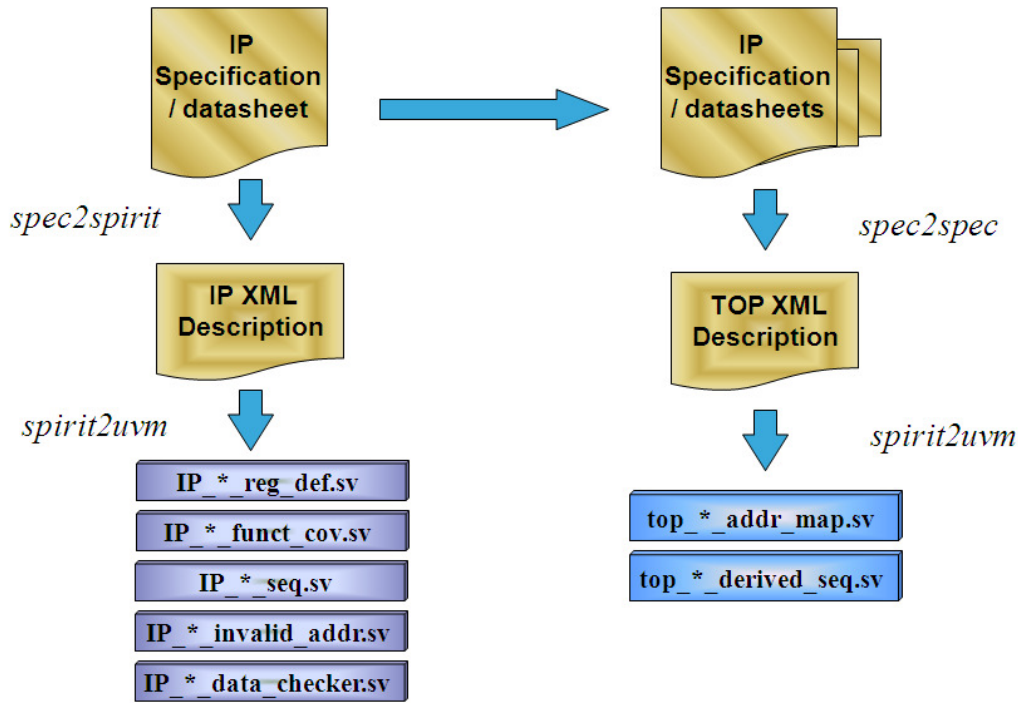

Figure. 10 IP-XACT Flow for generating IP and SoC level Verification Environment files

## 2.5 Register Verification

Standard RGM Register and Memory model is used for efficient register and memory verification. It contains built-in mechanisms with predefined types for efficient modeling. This is used in conjunction with the register interface UVC, so that whenever the IP/SoC registers are read/written, the associated RGM UVC pre-defined registers are also updated and IP/SoC register contents will be verified by a self-checking scheme.

## 2.6 Bit Accurate Verification

For the purpose of data checking, the System Verilog UVM based verification environment integrates the IP/ISP level python reference model. For control IP's, System Verilog scoreboard is written. Output of python model is compared with the output of the RTL in data checkers which are part of System Verilog UVM based verification environment.

## 3. CONCLUSIONS

We presented a Generic System Verilog UVM based reusable verification environment for verification of imaging IPs/SoCs both with Host BFM and actual Core using Incisive Software Extension (ISX) and Virtual Register Interface (VRI)/Verification Abstraction Layer (VAL) approaches. IP-XACT based tools are used for the automatic generation of IP/SoC dependent system verilog files. This verification environment has been started to be used for the verification of imaging IPs/SoCs. As compared to earlier methodologies, where verification flows were disjoint at IP, SoC and Validation level which was

- Increasing the verification effort and cost

- Increasing the maintenance of multiple verification environments

The proposed methodology helped in overcoming these drawbacks and helped in saving verification cost and effort.

Although this environment is developed for imaging IPs/SoCs. Same concept can be extended for non imaging IPs/SoCs. We are currently working on Formal Verification Methodology and Low-power simulation flow.

## ACKNOWLEDGEMENTS

The authors would like to thank management and team members of Imaging Division, STMicroelectronics; Faculty members and peer scholars of JBS, Jaypee Institute of Information Technology University and also cadence team for their support and guidance.

## Authors

**Abhishek Jain, Technical Manager, STMicroelectronics Pvt. Ltd.**
**Research Scholar, JBS, Jaypee Institute of Information Technology, Noida, India.**

Email: ajain_design@yahoo.co.in;
        abhishek-mmc.jain@st.com

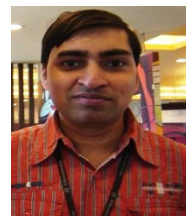

Abhishek Jain has more than 10 years of experience in Industry. He is responsible for Functional Verification Flow in Imaging Division of STMicroelectronics. He has done PGDBA in Operations Management from Symbiosis, M.Tech in Computer Science from IETE and M.Sc. (Electronics) from University of Delhi. His main area of Interest is Project Management, Advanced Functional Verification Technologies and System Design and Verification especially UVM based Verification, Emulation/Acceleration and Virtual System Platform. Currently he is doing Research in Advanced Verification Methodologies for Process improvement and control of Projects in Semiconductor Sector. Abhishek Jain is a member of IETE (MIETE).

**Dr. Hima Gupta, Associate Professor, Jaypee Business School (A constituent of Jaypee Institute of Information Technology University), A – 10, Sector-62, Noida, 201 307 India**.

Email: hima_gupta2001@yahoo.com

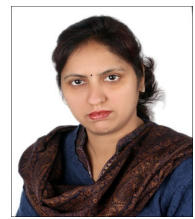

Dr. Hima has worked with LNJ Bhilwara Group & Bakshi Group of Companies for 5 yrs. and has been teaching for last 10 years as Faculty in reputed Business Schools. She also worked as Project Officer with NITRA and ATIRA at Ahmedabad for 5 years.
She has published several research papers in National & International journals.

**Giuseppe BONANNO, CAD Tools and Methodology Group Manager, STMicroelectronics Pvt. Ltd.**

Email: giuseppe.bonanno@st.com

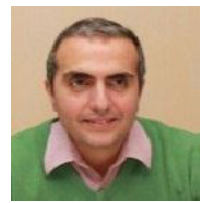

Degree in Computer Science at Milan State University (Italy) in 1989. Worked as CAD Engineer for Bull HN Italy (1989-1997) on the development of place and route tools for PCBs and then on ASIC hardware emulation. Joined STMicroelectronics in France in 1997 where he has worked on CAD support in different domains: FPGA prototyping, DFT, Functional Verification. Since 2006 he is managing the CAD Tools and methodology team of Imaging Division with the mission of developing and supporting Design tools and flows for modeling, IP and SOC design, functional verification and physical implementation.

**Ajay Goyal, Senior Sales Technical Leader, Cadence Design System Pvt. Ltd.**

Email: gajay@cadence.com

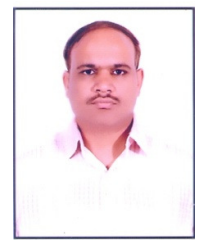

Ajay has more than 12 years of experience in EDA Industry. He has done MDBA from Symbiosis Institute of Management Studies (2002) and Bachelor of Engineering from MPIET, Nagpur University (1999). His main area of interest is Advanced Verification and System Design and Verification especially UVM based Verification, High Level Synthesis, Emulation/Acceleration and Virtual System Platform.